\documentclass[twocolumn,showpacs,prb,aps,amsmath,amssymb,superscriptaddress]{revtex4}

 \usepackage{amsmath}
 \usepackage{graphicx}
 \usepackage{rotating}

 \def\bbbc{{\mathchoice {\setbox0=\hbox{$\displaystyle\rm C$}\hbox{\hbox
 to0pt{\kern0.4\wd0\vrule height0.9\ht0\hss}\box0}}
 {\setbox0=\hbox{$\textstyle\rm C$}\hbox{\hbox
 to0pt{\kern0.4\wd0\vrule height0.9\ht0\hss}\box0}}
 {\setbox0=\hbox{$\scriptstyle\rm C$}\hbox{\hbox
 to0pt{\kern0.4\wd0\vrule height0.9\ht0\hss}\box0}}
 {\setbox0=\hbox{$\scriptscriptstyle\rm C$}\hbox{\hbox
 to0pt{\kern0.4\wd0\vrule height0.9\ht0\hss}\box0}}}}

 \usepackage{color}
 \usepackage{here}
 \definecolor{DarkBlue}{rgb}{0.1,0.1,0.5}
 \definecolor{Red}{rgb}{0.9,0.0,0.1}
 \definecolor{Green}{rgb}{0.0,0.99,0.0}

 \newcommand{\beq}{\begin{eqnarray}}
 \newcommand{\eeq}{\end{eqnarray}}

 \newcommand{\bk}{{\bf k}}
 \newcommand{\bQ}{\bf Q}

 \newcommand{\beqa}{\begin{eqnarray}}
 \newcommand{\eeqa}{\end{eqnarray}}

 \newcommand{\bb}{\bibitem}

\begin{document}

 \title{Incommensurate spin resonance in URu$_2$Si$_2$.}

 \author{A. V. Balatsky}
 \affiliation{Theoretical Division and Center for Integrated
 Nanotechnologies, Los Alamos National Laboratory, Los Alamos, New
 Mexico 87545, USA}

 \author{A. Chantis}
 \affiliation{Theoretical Division,   Los Alamos National Laboratory,
 Los Alamos, New Mexico 87545, USA}

 \author{Hari P. Dahal}
 \affiliation{Theoretical Division, Los Alamos National Laboratory,
 Los Alamos, New Mexico 87545, USA}

 \author{David Parker}
 \affiliation{U.S. Naval Research Laboratory,  4555 Overlook Ave. SW, Washington, DC 20375, USA}

 \author{J.X Zhu}
 \affiliation{Theoretical Division, Los Alamos National Laboratory,
 Los Alamos, New Mexico 87545, USA}

 \date{Printed \today }

 \begin{abstract}
 The nature of the hidden order (HO) in $URu_2Si_2$ below
 $T_{HO} = 17.5K$ has been a puzzle for a long time. Neutron scattering studies
 of this material reveal a rich spin dynamics.    We focus on inelastic neutron
 scattering in $URu_2Si_2$ and
 argue that observed  gap in the fermion spectrum naturally leads to the spin
 feature observed  at energies $\omega_{res} = 4-6
 meV$ at momenta
  at $\bQ^* = (1\pm 0.4, 0,0)$.  We
  discuss how spin features seen in $URu_2Si_2$ can indeed be thought
of in terms of {\em spin resonance} that develops in HO state and
  is {\em not related} to superconducting transition at 1.5K.
  In our analysis we assume that the HO gap is due to a particle-hole condensate that
 connects nested parts of the Fermi surface with nesting vector $\bf{Q}^* $.
 Within this approach we can predicted the behavior of the spin susceptibility at $\bQ^*$ and find it to be
 is strikingly similar to
 the phenomenology of resonance peaks in high-T$_c$ and heavy fermion
 superconductors. The energy of the resonance peak scales with $T_{HO}$
 $\omega_{res}  \simeq 4 k_BT_{HO}$. We discuss observable
 consequences spin resonance will have on neutron scattering and
 local density of states. Moreover,  we argue how establishment of spin resonance in $URu_2Si_2$  and
 better characterization
 of susceptibility,
 temperature, pressure and Rh doping
 dependence would elucidate the nature of
 the HO.
 \end{abstract}
 \pacs{Pacs Numbers: }

 \maketitle

 \vspace*{-0.4cm}

 \columnseprule 0pt

 \narrowtext \vspace*{-0.5cm}

 \section{Introduction}

 The problem of the nature of the hidden order below $T_{HO}=17K$ and
 the superconducting order below $T_c = 1.5K$ in URu$_{2}$Si$_{2}$
 has perplexed the condensed matter physics community for over two
 decades \cite{palstra}.

 The heavy-fermion (HF) superconductor URu$_{2}$Si$_{2}$ exhibits an
 order of an unknown origin which sets in at T$_{HO}=17.5 K$.
 Thermodynamic measurements \cite{maple} revealed a rather large jump
 of approximately 300 mJ/mol-K$^2$ in the linear specific heat
 coefficient $\gamma$ at 17.5 K. This material contains a linear
 specific heat coefficient $\gamma$ measured at 70-180
 mJ/mol-K$^{2}$, placing it as a moderately HF material. Below
 T$_{HO}$, the specific heat follows an exponentially activated
 behavior $\exp(-\Delta/T))$ with $\Delta$ estimated at 148 K. This
 gap also appears in optical measurements and vacuum tunneling and is
 comparable to that observed in inelastic neutron scattering
 experiments. Anomalies in the DC resistivity, Hall coefficient,
 thermal expansion and linear and nonlinear susceptibilities are also
 seen at T$_{HO}$, suggesting a substantial reordering of the
 conduction electrons. Neutron-scattering experiments \cite{maple,broholm}
 found an antiferromagnetic order below 17.5 K but with a staggered
 magnetization of only 0.03 $\mu_{B}$ per U atom, which is far too
 small to account for the observed specific heat anomaly. This
 anomaly correspond to an unobserved order which is therefore termed
 "hidden". Yet there are physical fields that clearly destroy hidden
 order. It is believed to be destroyed by an applied magnetic field
 of $\sim$ 40 T, suggesting a possible magnetic origin, but in
 \cite{santini} it was shown that there are two distinct
 field-independent energy scales, with opposite tendencies with
 magnetic field.  Therefore, any magnetic origin of this order must
 not couple directly to field in the same manner as the small AFM
 order. The application of pressure \cite{amitsuka} and Rh doping
 also suppress the HO \cite{amitsuka,curro}.

 Along with the determination of the experimental facts, there
 have been many theoretical attempts to understand this hidden order.
 Theories proposed include spin density waves of either
 unconventional or higher angular momentum character \cite{ramirez,
 ikeda}, orbital antiferromagnetism \cite{chandra}, staggered
 quadrupolar order \cite{santini1} and Jahn-Teller distortions
 \cite{kasuya}, multispin correlated order \cite{barzykin}, AFM
 states with anomalous g factors \cite{sikkema,ikeda}, valence
 admixture \cite{barzykin2}, octupole order \cite{hanzawa} and
 helicity order \cite{varma}. Determining the
 hidden order is complicated by possible phase separation into
 a magnetic moment phase and regions of hidden order, as argued by
 \cite{amitsuka}. To date no theory has shown conclusive agreement
 with the above experimental facts, and there exists no consensus
 as to the origin of the hidden order.

 Recently, \cite{wiebe}, Wiebe et al conducted an inelastic neutron
 scattering (INS) study of URu$_{2}$Si$_{2}$, in conjunction with
 specific-heat measurements above and below the 17.5 K onset
 temperature. Wiebe et al found that above the ordering temperature
 $T_{HO}$, gapless (with velocity $\sim v_{F}$) spin wave excitations
 centered on incommensurate wavevectors ${\bQ^*}=(1\pm0.4,0,0)$ appeared.
 But below this temperature these excitations were gapped, with
 an approximate gap at 1.5 K of 4-6 meV. Wiebe also estimated the
 specific heat coefficient of these gapless excitations and found a
 fair agreement with the experimental value. It was concluded that
 the reduction in specific heat below T$_{HO}$ resulted from the
 gapping of these spin-wave excitations; however, the order parameter
 responsible for this gapping remained indeterminate.

 The effect of opening a HO gap on spin excitations appears
 remarkably similar to the phenomenon of {\em spin resonance} in INS,
 seen in the superconducting state in cuprate
 materials,\cite{demler_rmp}
 and in the CeCoIn$_{5}$ superconductor \cite{stock}. For example, in the
 cuprates this resonance in the susceptibility $\chi({\bf q},\omega)$ is
 centered at the commensurate wavevector ${\bf q}$=($\pi,\pi$), and can be
 interpreted \cite{demler_rmp} as a bosonic mode transferring
 ${\bf q}$=($\pi,\pi$) from the neutron to the Cooper pair. For completeness of the discussion we also
 point the case of $Sr_2RuO_4$ where resonance was predicted but not observed to date \cite{srresonance}.
 One might
 therefore expect that a similar effect of gapping on the spin excitations can
 occur in a state with hidden order, even if the exact nature of HO
 is not yet settled.

 In this paper we propose that URu$_{2}$Si$_{2}$ should exhibit an
 {\em incommensurate spin resonance} based on an analogy with the
 inelastic neutron scattering resonance observed at 41 meV in the
 cuprates \cite{demler_rmp}. We argue that:

 1) The observation by Wiebe et al \cite{wiebe} of the substantial
 changes of spin susceptibility below and above $T_{HO}$ at an {\em
 incommensurate} momentum is indicative of the gapping of spin
 excitations due to the gapping of the electronic spectrum below
 $T_{HO}$. We have developed a mictroscopic theory of the
 spin susceptibility, outlined below. This theory
  is based on the estimate of changes in susceptibiltiy due to
 gap in fermionic spectrum. The gap in spin susceptibilty we
 estimate to be twice as large as a gap in single spin excitation as explained below.
 As a result we estimate $\omega_{res} \simeq 4 k_B T_{HO}$
 opening  that allows us to estimate the energy of the spin resonance to be in the
 range $\omega_{res}= 4-6meV $ and the momentum to be ${\bf Q}^*
 =(1\pm0.4,0,0)$. Changes in spin susceptibility
 due to the HO gap $\Delta_{Q^*}$ will naturally change spin excitation spectum. Given
 the mean field character of the HO
 gap opening as seen in the specific heat data, we expect  that the
 intensity of the resonance scales as $|\Delta_{\bQ^*}|^2 \sim
 (T_{HO}-T)$ below $T_{HO}$.

 2) Multiple orders were proposed as an explanation of HO.
 We argue that the experimental observations are consistent with a specific particle-hole
 order that has a finite {\em incommensurate} momentum
 ${\bQ^*}=(1\pm0.4,0,0)$ (and related by $k_x \leftrightarrow k_y$
 permutation) and leads to a gap in the spectrum $\Delta_{\bQ^*}$.
 The exact nature of this hidden order is likely be a hybridization gap
 $\Delta_{\bQ^*}$ that opens up due to the nesting of different parts
 of Fermi surface separated by ${\bQ^*}$. For our analysis of the
 spin susceptibility we focus on terms of second order in
 $\Delta_{\bQ^*}$ that would contribute to the spin susceptibility
 and therefore we do not need to know the exact details of the HO.
 Nevertheless our conclusion is that the data on INS and specific
 heat are consistent with the particle hole excitation being gapped
 below $T_{HO}$. Recent neutron scattering work by Janik et.al.
 \cite{janik} and theory proposal by Oppeneer \cite{oppenner}
 does point to the nesting phenomenon as a possible
 source of HO and is consistent with our proposal.

 3) The HO leads to spectral weight changes that produce a peak in
 the spin susceptibility which we call a {\em spin resonance} with
 energy $\omega_{res} = 4-6 meV$ at momentum ${\bQ^*}$. In the
 previous cases where a resonance peak has been seen in the ordered
 state, opening up a partial gap at the Fermi surface, this resonance
 peak has been observed at commensurate momenta. We point
that complicated spin dynamics  that is affected by the
 HO, in addition to already established spin gapping,  should exhibit a
  phenomena of {\em spin resonance peak } in $URu_2Si_2$. Main difference with respect
  to previous discussion on spin resonance is that this resonance  occurs at the
incommensurate momentum ${\bQ^*}$ in the nonsupercoducting state.

 To support our claim about fermion spectrum gapping, we will
 provide fits to the specific heat based on a mean field gap in the
 spectrum with the ratio $\Delta_{\bQ^*}/k_bT_c = 2.5$ that give a
 reasonably good fit to the data. We also address the density of states
 that can be measured by a scanning probe as another observable that
 might reveal the existence of an energy feature at $\omega_{res}$.

We present arguments  that naturally lead to the prediction of the spin resonance in $U
 Ru_2 Si_2$ in Sec II. Then we discuss observables such as the
 specific heat and the local density of states due to this resonance
 in Sec III. We conclude with a discussion section.

 \section{Spin Resonance in $U Ru_2 Si_2$}

 In the cuprates, the resonance in
 the susceptibility $\chi({\bf q},\omega)$ is centered at the commensurate
 wavevector ${\bf q}$=($\pi,\pi$), and can be interpreted \cite{demler_rmp}
 as a bosonic mode transferring $\bf q$=($\pi,\pi$) from the neutron to the
 Cooper pair. The energy of this resonance is independent of
 temperature, while its intensity depends strongly on temperature and
 vanishes at T$_{c}$. Within the SO(5) theory \cite{demler_rmp}
 linking superconductivity and magnetism in the cuprates, an
 excitation bearing these properties can arise naturally in the
 particle-particle superconducting channel, and leads to a resonant
 susceptibility $\chi(({\bf q}=\pi,\pi),\omega) \propto
 \Delta^{2}/(\omega-\omega_{res}+i\Gamma)$, where $\Delta$ is the
 superconducting order parameter and $\Gamma$ is a damping constant.
 This resonance peak appears only below T$_{c}$ because it is only
 below this temperature that the mixing of electrons and holes that
 occurs in the superconducting state allows coupling of magnetic
 excitations via particle-hole and particle-particle channel
 coupling. In the cuprates, this interaction is active within the
 superconducting particle-particle channel, but as we shall see it
 can be extended under suitable conditions to the particle-hole
 channel, leading to a similar result. In this case, however, the
 resonance occurs at an incommensurate wavevector, putting
 constraints on the origin of this resonance.

 In a more recently investigated case of CeCoIn$_{5}$, a similar
 resonance \cite{stock} is seen at $(\pi,\pi,\pi)$ and has been
 interpreted as evidence for d-wave symmetry. On the other hand, a spin
 resonance has been observed in the pnictide superconductor
 $Ba_{0.6}K_{0.4}Fe_2As_2$  where the pairing symmetry could be different
 \cite{pnictide}.

 We point out here that conflicting opinions on the possible origin of
 resonance peak exist. In particular,  alternative explanations of
 the resonance peak, including nonsuperconducting and purely magnetic
 commensurate response of incommensurate magnets,  also have been
 discussed \cite{batista,chubukov_prl}.  The relevance for the present
 discussion is that we do not see a need to have a superconducting
 referencce state as a prerequisite for spin resonance. The gapping of
 the spectrum is essential but the gap does not have to be
 superconducting. This is an important difference we stress: most of
 the cases of spin resonance were discussed with regards to superconductors. We do
 not imply here that URu$_{2}$Si$_{2}$ has superconducting
 correlations in the HO phase.

 \subsection{Spin Susceptibility}

 The suggestions of the previous section quickly lead to another
 option for connecting the formation of the hidden order with the
 spin dynamics. We propose a relatively simple explanation,
 consistent with the spin/hidden order coexistence, namely that a
 resonance peak in the susceptibility
 $\chi({\bQ^*},\omega=\omega_{res})$ appears as a result of the
 appearance of a particle-hole condensate, although more complex than
 the usual density-wave condensate. In particular, we  argue that the
 Fermi surface geometry, as depicted in Figure 1, is such as to allow
 an incommensurate nesting between the central $\Gamma$ Fermi surface
 pocket and the pocket, separated by $\bQ^*$. This nesting is not
 complete and would require a strong interaction to produce an
 instability. This fact is in accord with our observation that we
 need to use a strong coupling version of mean field specific heat to
 fit specific heat data, see below.
 \begin{figure}[h!]
 \includegraphics[width=8cm,angle=0]{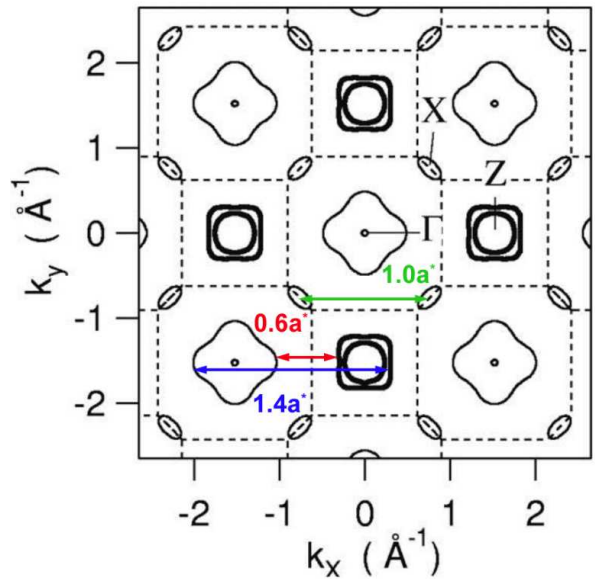}
 \caption{A depiction of the calculated Fermi surface geometry of
  URu$_{2}$Si$_{2}$, taken from \cite{janik}, with potential $ {\bQ^{*}}$ nesting
  vectors indicated
 entered at the corner of the Brillouin zone. However, this nesting
 is between two bands of with the same spin,  so that there is little or no
  magnetic signal and the order is "hidden".}
 \end{figure}

 We start with the calculation of spin-spin susceptibility, assuming
 that the particle hole ordering gaps the FS, which is nested with momentum
 ${\bQ^*}$. Assuming that the gap opens up below $T_{HO}$ we will argue
 that the change in susceptibility will have a term that is
 proportional to $\Delta^2_{\bQ^*}$. As such this second order
 correction will occur regardless of the detailed nature of the HO.
 Similar second order terms in the spin susceptibility for
 superconducting gap were argued for in earlier work\cite{demler_rmp}.

  We begin with the spin-spin susceptibility of itinerant electrons in $U
 Ru_2Si_2$ at T=0 signatures are seen in zz component $\chi^{zz}({\bQ^*},t) = i \langle T
 S^z({\bQ^*},t)S^z(-{\bQ^*},0)\rangle$ : \beqa \chi^{zz}(Q^*,\omega)=
 i\sum_{\bk\bk'} \int \langle T c^\dag_{\bk'+\bQ^*,\mu}(t)
 \sigma^z_{\mu,\nu}c_{\bk',\nu}(t) \nonumber\\
 c^\dag_{\bk-\bQ^*,\alpha}(0)
 \sigma^z_{\alpha
 \beta}c_{\bk,\beta}(0)
 \rangle e^{i\omega t} dt \nonumber\\
 = -i\sum_{\bk\bk'} \int \langle
 T c^\dag_{\bk'+\bQ^*,\mu}(t) c_{\bk,\beta}(0) \rangle
 \nonumber\\
 \langle T c^\dag_{\bk-\bQ^*,\alpha}(0) c_{\bk',\nu}(t) \rangle
 e^{i\omega t} \sigma^{\mu \nu}\sigma^z_{\alpha \beta} dt .
 \label{EQ:spinzz} \eeqa
 To make the next step we introduce the
 anomalous Green's functions that capture the appearance of
 incommensurate order at $\bQ^*$: \beqa F_{\bk,\bQ^*}(\omega_1)=i
 \langle T c^\dag_{\bk-\bQ^*,\alpha}(0) c_{\bk,\nu}(t) \rangle
 \delta_{\alpha \nu}
 \nonumber\\
 =\frac{\Delta_{\bQ^*}}{\omega_1^2-E^2_{\bk,\bQ^*} +i\delta},
 \label{EQ:F} \eeqa
 function $F$ describes particle hole density
 order that represents HO and is {\em nonmagnetic}. This order
 partially gaps excitations at the Fermi surface.  $F$ relates
 regions of the Fermi surface that are connected by nesting vector
 $\bQ^*$. We choose  it to have a typical mean-field form. As we have
 argued, with the focus on second order terms in $\Delta_{\bQ^*}$,
 the detailed structure of the propagators is not critical.  The same
 conclusions can be drawn from Ginzburg-Landau theory for the HO state.
 Hereafter we will ignore smooth terms in the susceptibility.  Then the
 susceptibility related to the appearance of anomalous order is \beqa
 \chi^{zz}({\bQ^*},\omega) =-i\sum_\bk \int
 F_{\bk,\bQ^*}(\omega_1) F_{-\bk,-\bQ^*}(\omega+\omega_1) d \omega_1.
 \label{EQ:chizz}
 \eeqa
 the integral in $\chi^{zz}(\bQ^*,\omega)$ can be written as,
 \beqa \chi({\bQ^*},\omega)=\Delta_{\bQ^*}\Delta_{-\bQ^*}\sum_{\bk}
 \frac{1}{2E_{\bk,\bQ^*}}
 \frac{1}{(E_{\bk,\bQ^*}+\omega)^2-E_{\bk, \bQ^*}^2} \\
 + \Delta_{\bQ^*}\Delta_{-\bQ^*}\sum_\bk \frac{1}{2E_{\bk, \bQ^*}}
 \frac{1}{(-\omega+E_{\bk, \bQ^*})^2-E_{\bk, \bQ^*}^2} \\
 =\Delta_{\bQ^*}\Delta_{-\bQ^*}\sum_\bk \frac{1}{E_{\bk, \bQ^*}}
 \frac{1}{\omega^2-4E_{\bk, \bQ^*}^2} \label{EQ:chizz2}
 \eeqa
 We took (see below) $N(E)=N(0)
 \frac{E}{\sqrt{E2-\Delta2_{\bQ^*}}}$ as is appropriate for a
 gapped spectrum, then \beqa
 \chi^{zz}({\bQ^*},\omega)=|\Delta_{\bQ^*}|^2
 \int \frac{1}{\sqrt{E^2-\Delta^2_{\bQ^*}}}\frac{1}{\omega^2-4E^2}dE.
 \label{EQ:chizz3}
 \eeqa
 Thus the susceptibility indeed acquires a term
 that scales quadratically with the HO gap. The details of the integral
 over energy in Eq(\ref{EQ:chizz3}) depend on the band structure.
 For any density of states that is smooth, simple analysis shows that
 for $\omega << \Delta$, $\chi^{zz}({\bQ^*},\omega) \propto
 |\Delta_{\bQ^*}|^2 \omega^2$, and for $\omega  \Delta_{\bQ^*}$,
 $\chi^{zz}({\bQ^*},\omega) \propto
 \frac{|\Delta_{\bQ^*}|^2}{\omega^2}$, with the crossover at $\omega
 \sim |\Delta_{\bQ^*}|$. We therefore immediately conclude that there
 is a resonance contribution to spin susceptibility $ \sim
 \Delta^2_{\bQ^*}$ and that contribution will have a peak at $\omega
 \sim \Delta_{\bQ^*}$.

Finally, we give an argument on why the spin resonance energy
$\omega_{res} = 4 k_B T_{HO} \sim 4-6 meV $. For any collective many
body state that develops full or partial gap in the mean field
transition at transition temperature $T_C$ the single particle gap
at low T will be on the order of \beqa \Delta_{qp} \geq 1.75 k_BT_C
\label{EQ:Gapcoeff1}
 \eeqa The single particle gap for any itinerant
system would lead to a spin gap on the order of {\em twice} the
single particle gap: \beqa \Delta_{spin} = 2 \Delta_{qp} = 3.5 k_B
T_C \label{EQ:Gapcoeff2} \eeqa
 Typical example is the density
wave where the gap opening suppresses low energy susceptibility and
opens up at least partial spin gap. For materials where strong
coupling effects are important, $URu_2Si_2$ is certainly one of
them, the typical single particle gap is larger with respect to weak
coupling coefficient of $1.75$. We therefore would expect that in
strong coupling systems where spin and charge interactions are
strong, the coefficient in Eq.(\ref{EQ:Gapcoeff2}) will be larger
and can reach value of $4-5$ and possibly higher. Spin resonance
energy reflects spin spectral weight redistribution  will occur at
energies on the order of $\Delta_{spin}$. So the generic relation
between $\omega_{res} = 4-5 k_B T_{HO}$ holds for $URu_2Si_2$ as
well because transition is demonstrably close to mean field with
single particle gap.

  We thus proved the points 1) and 2) we made in
 the Introduction.

 \section{ Experimental consequences}

 We now focus on the experimental observables  that can be used to test
 the prediction of a resonance peak in $URu_2Si_2$. We will consider
 neutron scattering and local density of states features. In addition
 we will address electronic specific heat features due to HO gap to
 illustrate that we can achieve a reasonable fit using a simple mean
 field description.

 \subsection{Inelastic Neutron Scattering}

 We expect a resonance peak with $\omega_{res}=4-6meV$ should appear
 in INS below T$_{HO}$, and the intensity of this peak should
 increase quasilinearly, as in prior work \cite{demler_rmp}, \beqa
 \delta\chi^{zz}(\textbf{q} = {\bQ^*}, \omega = \omega_{res}, T) \sim
 \Delta_{\bQ^*}^2 \sim |T-T_{HO}| \label{EQ:intensity1} \eeqa with
 decreasing temperature before saturating at low temperature ($<$ 0.6
 T$_{HO}$). This peak should be centered at the incommensurate
 wavevectors $(1\pm 0.4,0,0)$, \beq \delta\chi^{zz}(\textbf{q} ,
 \omega = \omega_{res}, T \ll T_{HO}) \sim
 \frac{\Delta_{\bQ^*}^2}{({\bf q}-{\bQ^*})^2 + \xi^{-2}}.
 \label{EQ:intensity2} \eeq

 The energy, momentum and temperature dependence of the resonance peak is
 illustrated in Fig(\ref{FIG:susceptibility})
 \begin{figure*}
 \includegraphics[width=.7\linewidth, angle=-90]{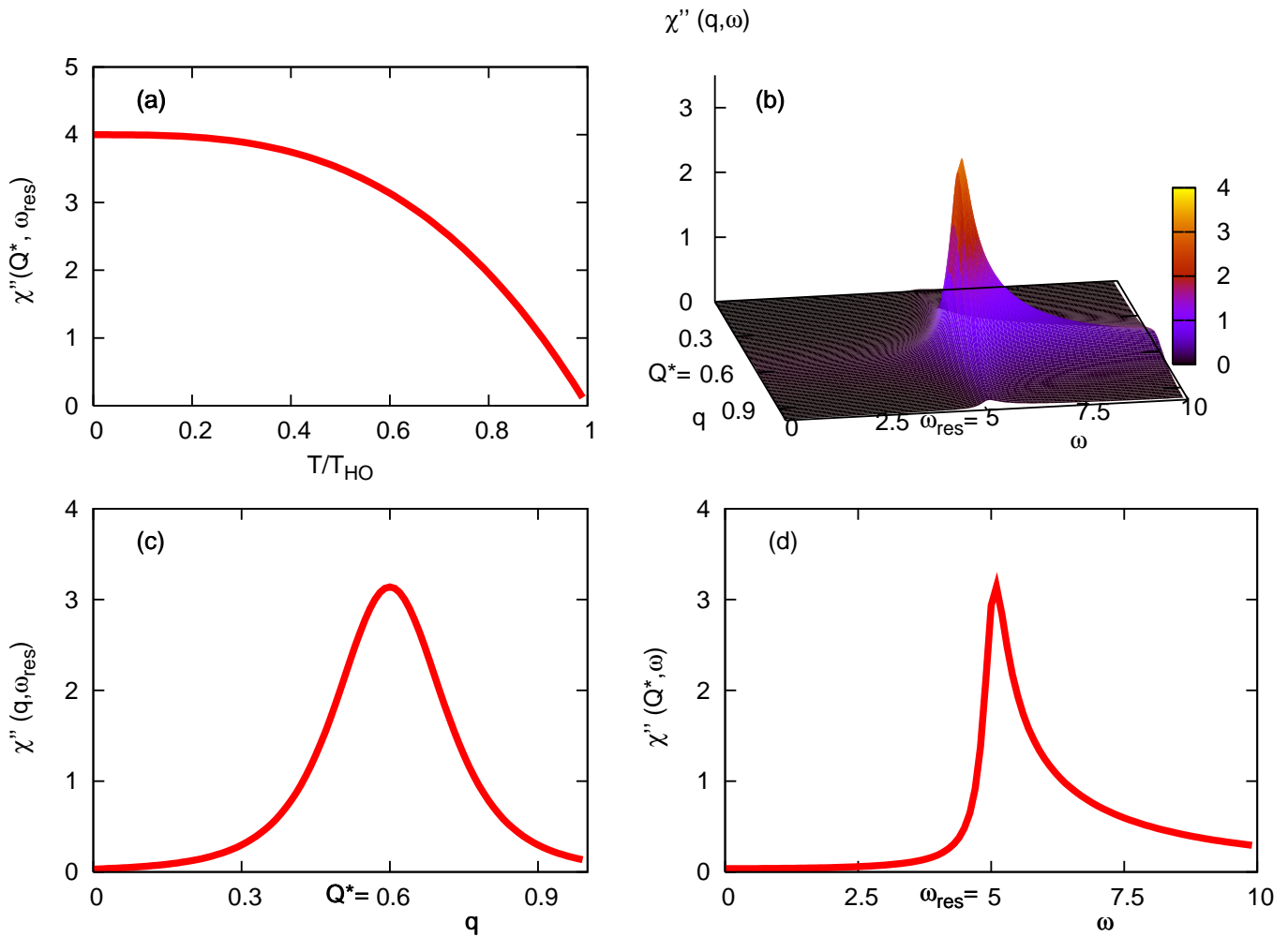}
 \caption{\label{FIG:susceptibility}(Color online) a) The spin
 susceptibility at the resonance momentum ${\bQ^*}$ and at the
 resonance energy $\omega_{res}$ is plotted as  a function of
 temperature normalized to the hidden order transition temperature,
 $\frac{T}{T_{HO}}$. The temperature dependence is shown to be
 determined by the temperature dependence of the "hidden-order" order
 parameter. b) The intensity plot of spin susceptibility near $\bQ^*$
 and $\omega_{res}$ is shown. It is clearly seen that the spectral
 weight of the susceptibility is transferred to the resonance
 momentum and the resonance energy. The spin susceptibility c) at the
 resonance energy as a function of the momentum and d) at the
 resonance momentum as a function of the energy are shown. }
 \label{FIG:susceptibility}
 \end{figure*}
 with a width $1/\xi$ depending on the microscopic details of the
 theory. From Ref[\cite{wiebe}] we estimate $\xi^{-1} \sim 0.1
 \frac{\pi}{a}$.

 From this analysis we would expect both intensity
 and resonance energy be temperature dependent. At temperatures below
 $T_{HO}$ resonance energy will evolve with temperature
 $\omega_{res} \sim \Delta_{Q*}$. In the same region intensity of
 resonance peak with change as a function of temperature $I({\bQ}*, \omega_{res},T) \sim
 \Delta^2_{Q*}$. Broholm et.al. \cite{broholm}  have shown that gap in the neutron scattering peak
 at $Q*$ does indeed depends on $T-T_{HO}$ in a mean field manner.
 They also argued that intensity of neutron scattering feature does
 increase below $T_{HO}$.
 Another mean to test dependence of resonance on HO
 gap is investigate effects of pressure or $Rh$ doping
 \cite{amitsuka,curro}. Resonance peak energy would be suppressed
 with $Rh$ doping. These estimates  can be tested experimentally.

 \subsection{Specific Heat}

 The gapping of fermions  on part of the Fermi surface directly results
 in the loss of entropy observed below T$_{HO}$, and we will
 demonstrate an excellent quantitative fit to the experimental
 specific heat data.

\begin{figure}[h!]
 \includegraphics[width=.9\linewidth]{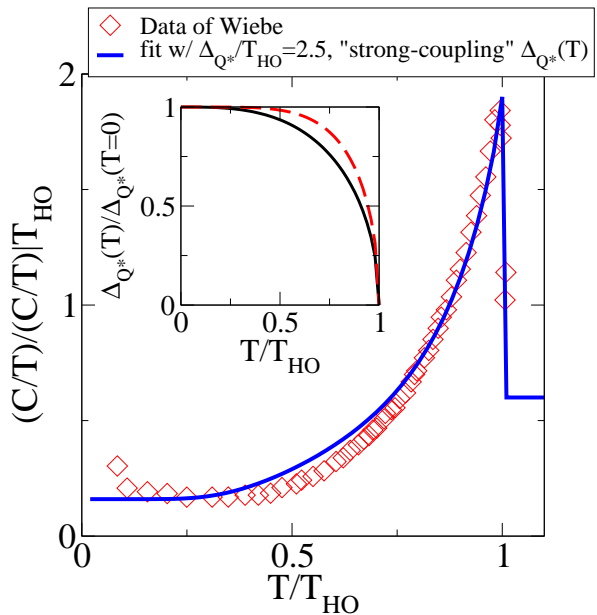}
 \caption{\label{fig:specificheat}(Color online) Shown are the
 results of a calculation of the electronic specific heat of
 $URu_2Si_2$ assuming that the Fermi surface is split into a gapped
 and ungapped region, with the order parameter taken at T=0 as
 2.5T$_{HO}$ $\simeq 3.75 meV$. For this calculation a temperature
 dependence of $\Delta_{\bf Q^*}(T)$ was assumed in which the gap
 develops below T$_{HO}$ more rapidly (inset, dashed line) than in
 canonical BCS theory (inset, solid line), as is often observed in
 strong-coupling superconductivity, but still maintaining mean-field
 character.  The gap $\Delta_{HO}$ used in the fit is assumed to be a
 FS averaged HO gap.} \label{FIG:Cplot3}
 \end{figure}

 In Figure \ref{FIG:Cplot3}) we plot the specific heat data of Wiebe
 \cite{wiebe}, and our fit, assuming a $\Delta_{\bQ*}/T_{c}$ ratio of
 2.5 and a ``strong-coupling'' temperature dependence of
 $\Delta_{\bQ*}(T)$. We work by analogy with the BCS theory of
 superconductivity, which shares many of the same expressions with
 this gapping of the Fermi surface \cite{gruner,schrieffer}.  In
 particular, the specific heat of the gapped portion of the system is
 given by \beq C(T) &=& 2k_{B}\alpha\beta2
 \sum_{\bk}f_{\bk}(1-f_{\bk})\left(E_{\bk}^2+\frac{\beta}{2}\frac{d\Delta^{2}}{d\beta}\right)
 \eeq where $E_{k}$ is the quasiparticle energy in the gapped state
 given by \beq E_{\bk} &=& \sqrt{\epsilon_{\bk}^2+\Delta_{\bf
 Q^{*}}^2} \eeq $\epsilon_{\bk}$ is the normal state dispersion,
 $\beta=1/k_{B}T$, and $\alpha$ is the gapped fraction of the Fermi
 surface. The jump in the specific heat at T$_{HO}$ is caused by the
 second term of the above equation.  The effect of this term is
 enhanced both by the $\Delta_{\bQ^{*}}(0)/T_{HO}$ value of 2.5
 exceeding the BCS weak-coupling value of 1.76 and by the assumed
 ``strong-coupling'' form of $\Delta_{\bf Q^*}(T)$, in which the
 quasiparticle gap develops more rapidly below T$_{HO}$ than in
 standard BCS theory.  Such a rapid gap opening is well-known from
 studies of the cuprates \cite{monthoux,eliashpaper}, and can occur
 due to the rapid suppression of bosonic excitations below T$_{HO}$.
 The gap still retains a square-root singularity at T$_{HO}$, and
 hence a mean-field, second-order phase transition at this
 temperature. Comparing the numbers from strong coupling theory with
 the data we see a reasonable agreement: $\Delta_{HO} \sim 4-6 meV$,
 $T_{HO} = 17K$ and $\Delta_{\bQ^{*}}(0)/T_{HO} \sim 2.3$.

 To the gapped specific heat must be added a term from the ungapped
 portion of the Fermi surface, given simply by $C_{n}=(1-\alpha)\gamma T$, with
 $\gamma$ the Sommerfeld specific heat coefficient.  For this
 calculation approximately 60 percent of the Fermi surface was assumed
 to be gapped.

 We have not included in the calculation the effects of the phonon
 specific heat or of the apparently correlation-induced rise in C/T at
 very low temperature; these effects have opposite temperature
 dependencies and
 are of
 comparable magnitude, so that the overall effect on the fit
 of neglecting these effects is expected to be small.

 \subsection{Local density of states}

 Here we present a simple qualitative argument that shows how the
 Local Density of States (LDOS) can be used to reveal the resonance
 peak. The energy of the resonance makes its observation relatively
 simple with a Scanning Tunneling Microscope (STM). The main feature
 that we focus on is the LDOS at the tunneling bias that reveals the
 energy gap in the electron spectrum in the range $2-4 meV$. We begin
 by assuming a typical ordered state self-energy
 \beq \Sigma({\bf
 k},\omega) &=& \frac{|\Delta_{\bf Q^{*}}|^2}{\omega-\epsilon_{{\bk}+{\bf
 Q}^{*}}} \eeq
 which can then be combined with Dyson's equation: \beq
 G(\omega,{\bf k}) &=& \frac{1}{\omega-\epsilon_{\bf k}-\Sigma({\bf
 k},\omega)+i\delta} \eeq Solving for the poles of the Green's
 function gives the quasiparticle dispersion relation as
 \beq \omega
 &=& \frac{\epsilon_{\bf k}+\epsilon_{\bf k+Q*} \pm
 \sqrt{(\epsilon_{\bf k}-\epsilon_{\bf k+Q*})^2+4|\Delta_{\bf
 Q*}|^2}}{2} \eeq
 which is the dispersion for a density-wave nested
 at ${\bf Q^{*}}$.  In particular, if ${\bf k}$ and $\bk +\bQ^*$ are
 on the gapped portion of the Fermi surface, we obtain a simple
 gapped spectrum
 \beq \omega = \pm \Delta_{\bf Q^*} \eeq
 The local density
 of states will depend to a certain extent on the details of the
 dispersion, which we have not attempted to model here.  A summation
 over the whole Fermi surface will lead to the finite DOS $N(\omega =
 0)$. In general the LDOS  will contain a feature at $E=\pm\Delta_{\bf
 Q^*}$ from the usual density-of-states relationship of a gapped
 spectrum, \beq N(\omega)=
 N_{0}\frac{\omega}{\sqrt{\omega^2-|\Delta_{\bf Q^*}|^2}} \eeq Such a
 feature should be readily observable by low-temperature STM for
 E=2-4 meV, although the effects of impurities and inhomogeneities
 will tend to broaden this peak.

 \section{Relation to Superconductors  that Exhibit Resonance Peak}

 \begin{figure*}
 \includegraphics[width=1\linewidth]{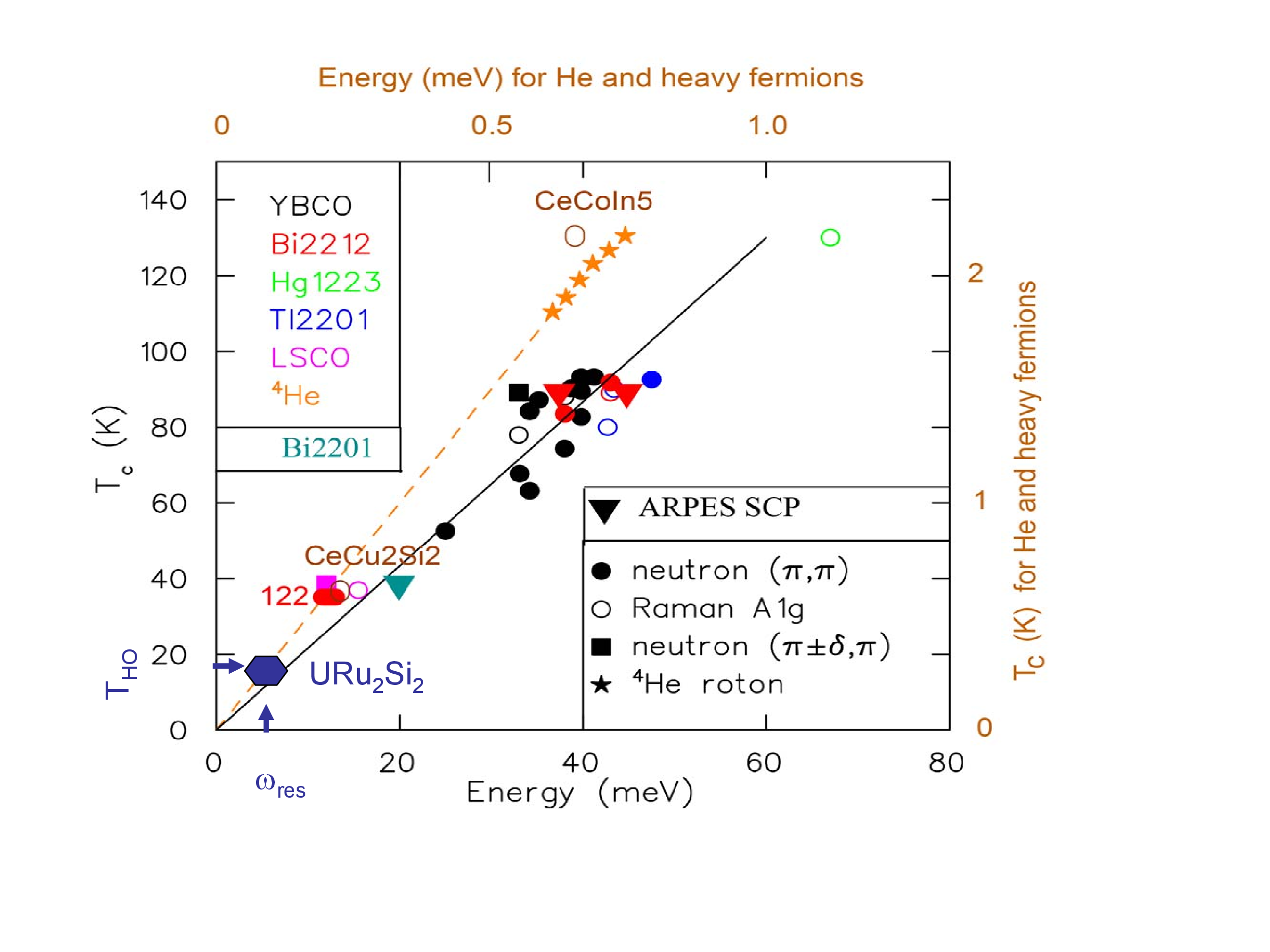}
 \caption{\label{FIG:resenergyplot}(Color online) The relation
 between the resonance energy $\omega_{res}$ and $T_c$ is shown for a
 variety of superconductors in this Uemura roton plot.  At a lower left corner we have added
 the point, indicated by arrows,  that marks HO relation between
 expected resonance peak and $T_{HO}$. The graph for superconductors and superfluid He is
taken from [\cite{uemura}].
 }
 \end{figure*}

 There is  an interesting correspondence between the energy of the resonance
 peak in $URu_2Si_2$ and in superconductors. Assuming that our
 prediction about the temperature dependence and the energy of the resonance peak
 is supported by experiment, we expect the resonance energy to be in
 the range $\omega_{res} = 4-6 meV$ and it to occur below $T_{HO}$.

 The relation between energy and critical temperature for HO phase is
 remarkably similar  to the relation between resonance energy and
 $T_c$ for unconventional  superconductors.
 For $URu_2Si_2$ HO state we find the ratio \beq \hbar \omega_{res}
 \simeq 4 k_BT_{HO} \label{EQ:resenergy2} \eeqa
 that is very similar to superconducting relation:
 \beqa \hbar \omega_{res} = 4 k_B T_c \label{EQ:resenergy} \eeqa
 We do not know the specific reason for
 this close correspondence other than the general observation that
 a gapped spectrum could also produce suppressed spectral weight in the spin susceptibility.

  Uemura [\cite{uemura}]
 noticed a universal scaling
between resonance energy and critical temperature for unconventional
super-conductors like high-Tc cuprates \cite{demler_rmp},
CeCoIn5\cite{stock} where it is seen at $(\pi,\pi,\pi)$ and in the
pnictide superconductor $Ba_{0.6}K_{0.4}Fe_2As_2$ \cite{pnictide}.
He proposed an analogy of resonance mode with rotons in superfluid
$^{4}$He using a plot shown in Fig.(\ref{FIG:resenergyplot}).  We
note that datum for HO phase is remarkably close to relation  for
superconductors and He, as demonstrated by a new point for
$URu_2Si_2$  added in Fig. \ref{FIG:resenergyplot}.    This analogy,
while appealing, can only go up to a point, since HO state in
$URu_2Si_2$ is non-superconducting and resonance feature is
incommensurate.

 \section{Discussion and Conclusion}

 In conclusion, we propose to search for the spin resonance in
 $URu_2Si_2$ at $\omega_{res} = 4-6 meV$ at the {\em incommensurate
 wavector} $Q^* = (1\pm0.4, 0,0)$. We expect that this spin resonance
 will set in at temperatures below the HO transition and the intensity of
 this peak will scale as $\sim \Delta^2_{HO} \sim (T_{HO}-T)$.

 The resonance peak is known to occur in the states with
 superconducting gap and results in the gapping of the electronic
 spectrum \cite{demler_rmp,stock,pnictide}. In the case of HO the gap
 $\Delta_{HO}$ results in the partially gapped electron spectrum.
 That appears to be a sufficient condition, as shown by Wiebe et al
 \cite{wiebe} to produce a gap in spin excitation spectrum.

There are few ways one can further experimentally test the predicted
relation between $T_{HO}$ and resonance energy with temperature,
impurity doping, pressure and with magnetic field. Resonance energy
$\omega_{res} \sim \Delta_{Q*}$ is a monotonic function of
temperature and $Rh$ doping. Similarly intensity of resonance peak
in critical region will scale as $\Delta^2_{Q*}$. Upon adding $Rh$
and $Th$ into $URu_{2-x}Rh_xSi_2$ one can suppress HO and respective
transition temperature and we would expect that the resonance energy
$\omega_{res}$ would track $T_{HO}(x)$. Similarly one can measure
changes in transition temperature and in resonance energy as a
function of pressure and magnetic field, if this is feasible.

The resonance discussed here is, to the best of our knowledge,   the
first case where the spin resonance occurs at an incommensurate
vector ${\bQ^*}$.

 The authors thank C. Batista, E. Bauer, C. Broholm,  P. Chandra, P. Coleman, N. Curro,  J.C Davis, J. Floquet, M.
 Graf, G. Kotliar, J. Lashley G. Luke, P. Oppeneer, F. Ronning, J. Sarrao, Y.J.  Uemura and C. Wiebe for
 useful discussions. This work was supported by US DOE at Los Alamos.
 D.P. is grateful to LANL and the T4 group for hospitality.

 \end{document}